\documentclass{moriond}

\def\JCAP{{\em JCAP}}
\def\JHEP{{\em JHEP}}

\def\be{\begin{equation}}
\def\ee{\end{equation}}
\def\bea{\begin{eqnarray}}
\def\eea{\end{eqnarray}}

\usepackage{amssymb}

\def\ba{\begin{eqnarray}}
\def\ea{\end{eqnarray}}
\def\mpl{M_{\rm Pl}}
\def\K{{\cal K}}
\def\mn{_{\mu \nu}}

\def\L{\mathcal{L}}

\def\g{\gamma}

\def\({\left(}
\def\){\right)}
\def\p{\partial}
\def\stu{St\"uckelberg}
\newcommand{\refeq}[1]{(\ref{#1})}
\newcommand{\eqref}{\refeq}
\def\T{T}
\def\P{\tilde \pi}
\def\Ein{\hat{\mathcal{E}}}

\newcommand{\Photo}{\includegraphics[height=35mm]{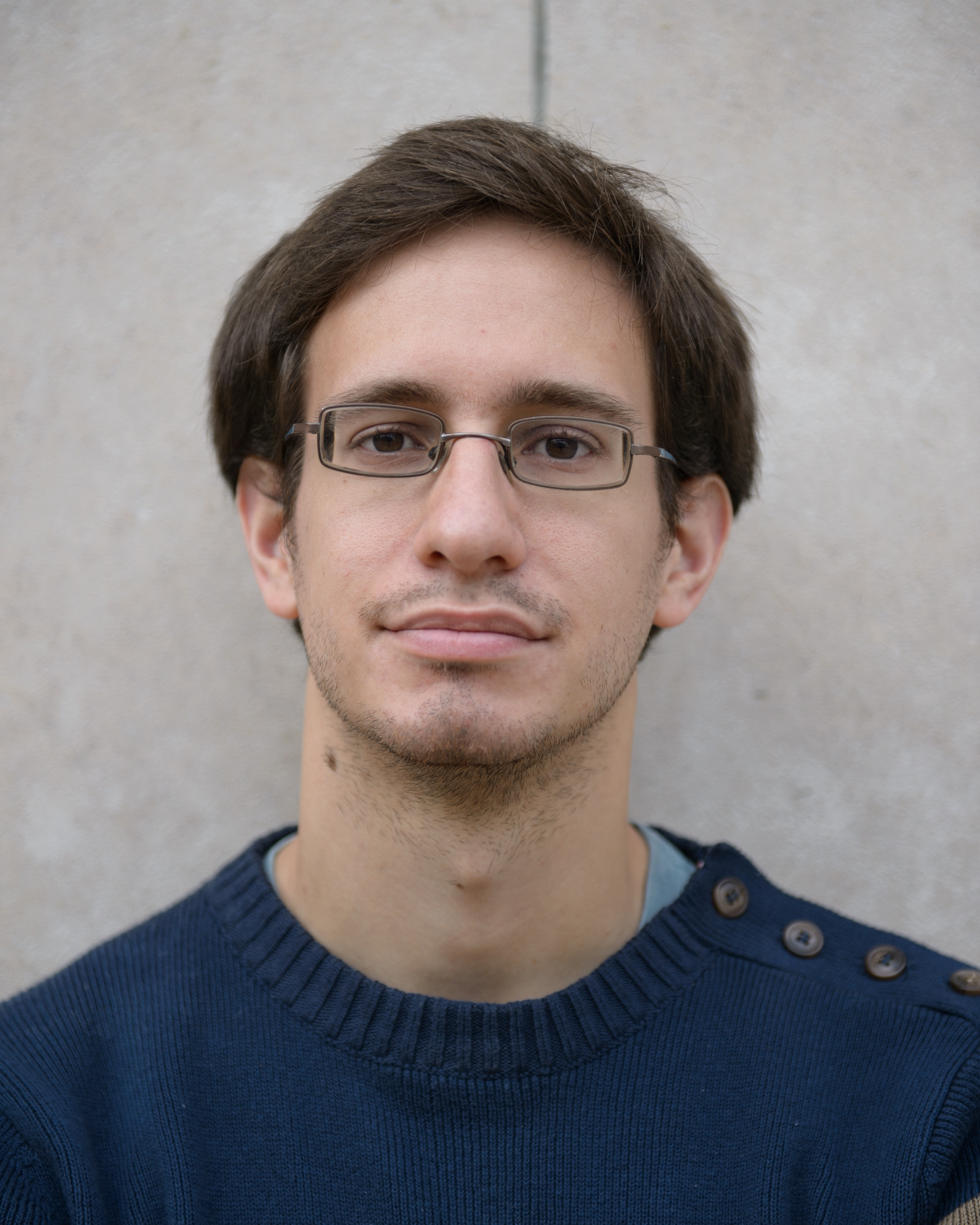}}

\begin{document}
\vspace*{4cm}
\title{ASPECTS OF MASSIVE GRAVITY}

\author{ SEBASTIEN RENAUX-PETEL }

\address{${\cal G} \mathbb{R} \varepsilon \mathbb{C} {\cal O}$ Institut d'Astrophysique de Paris, UMR 7095, CNRS,\\
Sorbonne Universit\'es et UPMC Univ Paris 6, 98 bis boulevard Arago, 75014 Paris, France.}

\maketitle\abstracts{
We report here on two works on Lorentz invariant massive gravity. In the first part, we derive the decoupling limit of massive gravity on de Sitter, relying on embedding de Sitter into an higher dimensional Minkowski spacetime. This enables us to identify the unique candidate for a partially massless gravity theory, in which only four degrees of freedom propagate, although further work showed that this property does not hold beyond the decoupling limit. In the second part, we study the fate of the Vainshtein mechanism in the minimal model of massive gravity, in which we show the limits of the often used assumptions of staticity and spherical symmetry.}

\section{Introduction}

Is it possible to give a mass to the graviton? While it can be motivated by a possible resolution of the old cosmological constant problem, this question is also theoretically interesting on its own right. It actually has a long and complex history, dating back from Fierz and Pauli \cite{Fierz:1939ix} in 1939, passing from important works in the 70s, to recent breakthroughs in the past few years (see the reviews \cite{Hinterbichler:2011tt,deRham:2014zqa}). In particular, de Rham, Gabadadze and Tolley recently identified the unique class of Lorentz-invariant massive gravity theories (henceforth dRGT) \cite{deRham:2010ik,deRham:2010kj} devoid of the so-called Boulware-Deser ghost, a fatal pathology that was present in all previous attempts. This class of theories, like any massive gravity theory, requires the introduction of a second metric beyond the ``standard'' one. This second, so-called reference metric, is usually chosen to be the one of Minkowski spacetime, but it can actually be arbitrary, and even dynamical \cite{Hassan:2011zd}. However, the physical content of such enlarged class of theories is largely unknown. In section \ref{deSitter}, we summarize a study of dRGT massive gravity on de Sitter spacetime \cite{deRham:2012kf}. This maximally symmetric set-up can be seen as the simplest extension of the theory around Minkowski, and is also motivated by cosmological reasons. Observational consistency of massive gravity theories requires that their additional degrees of freedom compared to General Relativity (GR) are hidden near dense sources, to conform for instance with gravity precision tests in the solar system. This non-trivial task is endorsed by the Vainshtein mechanism \cite{Vainshtein:1972sx,Babichev:2009jt,Babichev:2013usa}, in which non-linear effects render the new degrees of freedom strongly kinetically self-coupled, so that they almost do not propagate. In section \ref{Vainshtein}, we report on a study of the fate of the Vainshtein mechanism in the minimal model of massive gravity \cite{Renaux-Petel:2014pja}, in which we show the limits of the often used assumptions of staticity and spherical symmetry.

\section{The action of dRGT massive gravity}

The dRGT massive gravity action reads, in $d$ spacetime dimensions \cite{deRham:2010kj}:
\ba
\label{MG_lagrangian}
\mathcal{L}_{MG}=\frac{\mpl^{d-2}}{2}\sqrt{-g}\left(R-\frac{m^2}{4}\mathcal{U}(g)\right)\,,
\ea
where the first term is the Einstein-Hilbert action of GR, $m$ will be identified with the mass of the graviton, and the most general potential $\mathcal{U}$ is given by
\ba
\label{pot}
\mathcal{U}(g)&=&-4 \sum_{n=2}^d \alpha_n \L_{\rm der}^{(n)}(\K)\,.\\
{\rm with} \quad \L_{\rm der}^{(n)} (\K)&=&-\frac{1}{(d-n)!}  \mathcal{E}^{\alpha_1 \cdots \alpha_d} \mathcal{E}_{\beta_1 \cdots\beta_n \alpha_{n+1} \cdots \alpha_d}\
\K_{\alpha_1}^{\beta_1}\cdots \K_{\alpha_n}^{\beta_n}\,,
\ea
where $\mathcal{E}^{\alpha_1 \cdots \alpha_d}$ is the fully antisymmetric Levi-Cevita tensor and indices are raised and lowered using the metric $g\mn$. There, $\mathcal{K}^{\mu}_{\,\nu}$ is defined as $\mathcal{K}^{\mu}_{\,\nu}=\delta^{\mu}_{\nu}-\sqrt{g^{\mu\alpha}\gamma_{\alpha\nu}}$, where $\gamma_{\mn}$ is the reference metric, and we will only be concerned with cases in which $g_{\mu \nu}$ is close to $\gamma_{\mu \nu}$, so that $g^{\mu \alpha} \gamma_{\alpha \nu}$ is close to the identity matrix, and the matrix square root in $\K$ is well defined in perturbation theory. For definiteness we will always choose $\alpha_2=1$ while the other coefficients $\alpha_n$ are arbitrary.

\section{The decoupling limit of massive gravity on de Sitter}
\label{deSitter}

A theory of massive gravity free of the Boulware-Deser ghost propagates 5 degrees of freedom. Around Minkowski spacetime, this comprises two helicity-2 modes, like in GR, two helicity-1 modes and one helicity-0 mode. However, only around a maximally symmetric spacetime does it make sense to perform a helicity decomposition of a spin-2 field. Around an arbitrary reference metric, one no longer has a full Poincar\'e or equivalent group, and there is therefore no Poincar\'e representation to talk about. Since de Sitter (dS) spacetime is also a maximally symmetric manifold, the notion of a helicity decomposition around this spacetime is meaningful, but its identification requires additional work. The strategy we used is to embed $d$-dimensional dS into $d+1$-dimensional Minkowski spacetime. There, the identification of the various helicity modes relies on the well known \stu\, trick, in which the broken diffeomorphism invariance is restored while making the new degrees of freedom explicit \cite{ArkaniHamed:2002sp}. The subtlety then lies in their projection back into $d$-dimensional dS. Following this, we obtained the expression of the $d$-dimensional covariantized reference metric as
\be
\tilde \g\mn =\g_{\mu \nu}-S_{\mu \nu}-S_{\nu \mu}+ S_{\mu \alpha} \g^{\alpha \beta} S_{\nu \beta} +\frac{H^2}{1- H^2 V^2} \T_{\mu} \T_{\nu}\,,
\label{explicit-f}
\ee
with
\be
S_{\mu \nu}=\nabla_{\mu} V_{\nu}+\g_{\mu \nu}  \left(1-\sqrt{1-H^2 V^2 } \right)\,, \qquad \T_{\mu}= \frac 12 \p_\mu V^2 -  \sqrt{1-H^2 V^2 }\,  V_{\mu}
\ee
and where $V^2= \gamma_{\mu \nu} V^{\mu} V^{\nu}$, $H^2$ is proportional to the scalar curvature of dS spacetime $R=d(d-1)H^2$, and all the covariant derivatives are with respect to $\g_{\mu \nu}$. At this stage, we may split $V_\mu$ into $V_\mu = A_\mu +\p_\mu \P$, where at the linearized level, and in the decoupling limit that we will define below, $A_\mu$ describes the helicity-1 mode and is a vector field, while $\P$ is a scalar field that successfully encodes the helicity-0 mode. From the above expression, it is straightforward to deduce the structure of the linearized fluctuations, and to recover the so-called Higuchi bound, namely that one should have $m^2 > (d-2) H^2$ to ensure that all fields have a positive kinetic energy \cite{Higuchi:1986py}. 

Beyond the free theory, the decoupling limit is constructed in such a way as to concentrate on the interactions arising at the lowest energy scale, and and to disentangle them from the standard complications and non-linearities of GR. To achieve this, we simultaneously send $\mpl \to \infty$ (effectively keeping the helicity-2 modes $\tilde h_{\mu \nu}=g_{\mu \nu}-\gamma_{\mu \nu}$ linear) and $m \to 0$, in such a way as to keep finite both the canonically normalized fields and the lowest energy scale of interactions of the helicity-0 mode, \textit{i.e.}:
\ba
h\mn = \mpl^{(d-2)/2} \tilde h\mn \to {\rm finite}\,, \quad \pi =\Lambda^{(d+2)/2} \P \to {\rm finite}\,, \quad \Lambda \equiv (m^2\mpl^{(d-2)/2})^{2/(d+2)} \to {\rm finite}
\ea
To satisfy the Higichi bound as $m \to 0$, we also send $H$ to $0$, and we keep the ratio $\beta \equiv H^2/m^2$ finite so as not to recover the flat space limit. In this decoupling limit, the helicity-1 mode always arises quadratically and can hence be consistently set to zero, which we do in the following. After a long and complicated calculation, one finds the decoupling limit Lagrangian:
\ba
\L_{{\rm DL}}= -\frac14 \bar h^{\mu \nu} \Ein^{\alpha \beta}\mn \bar h_{\alpha \beta}
+\frac12 \bar h^{\mu \nu} \sum_{n=3}^{d-1}(\alpha_n+(n+1) \alpha_{n+1}) \frac{X^{(n)}\mn }{ \Lambda^{(d+2)(n-1)/2}}  +\sum_{n=2}^{d+1} c_n  \frac{\L^{(n)}_{\rm Gal}  }{ \Lambda^{(d+2)(n-2)/2} } \,,
\label{final}
\ea
where one partially diagonalized the action by use of the transformation
\ba
h\mn=\bar h\mn +\frac{2}{d-2} \pi \eta\mn -\frac{1+3 \alpha_3}{\Lambda^{(d+2)/2}}\p_\mu \pi\p_\nu \pi\,,
\ea
and the first term has the functional form of the linearized Einstein Hilbert action. The tensor $X^{(n)}\mn$ is a polynomial function of order $n$ of the covariant second derivative $\Pi_{\mu \nu}=\nabla_\mu \nabla_{\nu} \pi$, the $c_n$'s are $\beta$ and $\alpha_m$'s dependent coefficients, and 
\ba
\L_{\rm Gal}^{(n)} = (\p \pi)^2 \L^{(n-2)}_{\rm der}(\Pi)
\ea
are the so-called Galileon Lagrangians, with $\L_{\rm der}^{(0)}=1$. The decoupling limit is hence qualitatively very similar to that on Minkowski, and the appearance of the Galileon terms testifies that we correctly identified the helicity-0 mode $\pi$, whose equations of motion are manifestly second-order. 

When exploring this decoupling limit in more depth, we can unveil the existence of a very peculiar set of parameters: for the choice $\beta=1/(d-2)$, $\alpha_3=-\frac13 \frac{d-1}{d-2}$ and $\alpha_n=-\frac{1}{n} \alpha_{n-1}\, \, {\rm for} \, \, n \geq 4$, it turns out that the helicity-0 mode completely disappears from the decoupling limit Lagrangian! The corresponding model therefore represents the unique fully non-linear candidate theory of partially massless gravity, in which only $4$ degrees of freedom would propagate. This important identification prompted an important number of further studies \cite{Deser:2012qg,Hassan:2012gz,Deser:2013bs,Deser:2013uy,deRham:2013wv}, which established the reappearance of the helicity-0 mode beyond the decoupling limit.

\section{The Vainshtein mechanism beyond staticity and spherical symmetry}
\label{Vainshtein}

Any gravitational theory must conform with gravity precision tests in the solar system. In the context of massive gravity theories, the Vainshtein mechanism aims at screening the helicity 1- and 0-modes near dense sources to recover GR. In most studies, this amounts to establishing that in vacuum and static spherically symmetric (SSS) configurations, one can find a viable solution that interpolates between the Schwarzschild solution at sufficiently small radius from the source, and the expected Yukawa-type solution at large distances. Additionally, to gain some analytical insight, one often uses the decoupling limit approximation presented in the previous section. In this respect, the so-called minimal model is particularly interesting. It corresponds to a Minkowski reference metric and to the choice of parameters $\alpha_3=-1/3$ and $\alpha_4=1/12$ in the potential \refeq{pot}. For this particular choice, the decoupling limit Lagrangian \refeq{final} simply describes a free theory! That is, no interactions arise at the energy scale $\Lambda=(m^2 \mpl)^{1/3}$, contrary to all other choices of parameters (note that the helicity-0 mode does appear though, but only through its kinetic term). It can be tempting to infer from this absence of non-linear interactions at this scale that the Vainshtein mechanism is ineffective in this model \cite{Koyama:2011yg}. However, it only testifies that the decoupling limit, which aims at concentrating on the lowest energy interaction, has not been correctly identified at this particular point in parameter space. Considering SSS configurations, we actually showed that all interactions below the Planck mass identically vanish in the helicity-2 and -0 sectors \cite{Renaux-Petel:2014pja}. This tantalizing fact does point towards an absence of Vainshtein mechanism in this set up, but does not prove it. For this reason, we resorted to the exact equations of motion in the metric formalism. We then showed completely generally that in all vacuum SSS configurations, there exists an obstruction that precludes any recovery of General Relativity.  

While this could be seen as a proof that the minimal model is ruled out, we argue that this would be premature to reach this conclusion without further study. Indeed, we showed that in generic time-dependent or non-spherically symmetric configurations, interactions arbitrary close to the scale $\Lambda$ reappear! Although it is hard to reach conclusions solely on these facts, one can thus wonder whether the small degree of spherical symmetry breaking in the solar system can be enough to lead to a successful Vainshtein mechanism in the minimal model. More generally, while screening mechanisms have been mostly studied in static/stationary spherically symmetric situations up to now (see however ref \cite{Babichev:2011iz}), our analysis leads us to question whether the high degree of symmetry of these configurations might miss some important physical phenomena that arise in nature in realistic circumstances.

\section*{Acknowledgments}

This work was supported by French state funds managed by the ANR within the Investissements d'Avenir programme under reference ANR-11-IDEX-0004-02. I would like to thank the scientific committee of the Moriond conference for their financial support.

\end{document}